\pdfoutput=1
\documentclass{article}
\usepackage{graphicx, indentfirst, hyperref}
\usepackage[a4paper, margin = 3cm]{geometry}
\usepackage[numbers, sort&compress]{natbib}
\makeatletter
\let\mailmark\@fnsymbol
\makeatother

\newcommand*{\cf}{\emph{cf.}}
\newcommand*{\eg}{\emph{eg.}}
\newcommand*{\ie}{\emph{i.e.}}
\newcommand*{\prog}[1]{\emph{#1}}
\newcommand*{\figref}[1]{Figure \ref{fig:#1}}
\newcommand*{\secref}[1]{Section \ref{sec:#1}}
\let\thxmark\textsuperscript
\begin{document}
\title{Progress and outlook on advanced fly scans based on \prog{Mamba}}
\author{%
	Peng-Cheng Li\thxmark{1}, Cheng-Long Zhang\thxmark{1},
	Zong-Yang Yue\thxmark{1}, Xiao-Bao Deng\thxmark{1},\\
	Chun Li\thxmark{1}, Ai-Yu Zhou\thxmark{1,\mailmark{1}},
	Gang Li\thxmark{1}, Yu Liu\thxmark{1,\mailmark{1}}, Yi Zhang\thxmark{1,2}
}
\date{}
\maketitle
\begingroup
\renewcommand{\thefootnote}{\fnsymbol{footnote}}
\footnotetext[1]{\ %
	Correspondence e-mail:
	\texttt{zhouay@ihep.ac.cn}, \texttt{liuyu91@ihep.ac.cn}.%
}
\endgroup
\footnotetext[1]{\ %
	Institute of High Energy Physics, Chinese Academy of Sciences,
	Beijing 100049, People's Republic of China.%
}
\footnotetext[2]{\ %
	University of Chinese Academy of Sciences,
	Beijing 100049, People's Republic of China.%
}

\section*{Abstract}

Development related to PandABox-based fly scans is an important part of the
active work on \prog{Mamba}, the software framework for beamline experiments
at the High Energy Photon Source (HEPS); presented in this paper is the progress
of our development, and some outlook for advanced fly scans based on knowledge
learned during the process.  By treating fly scans as a collaboration between
a few loosely coupled subsystems -- motors / mechanics, detectors / data
processing, sequencer devices like PandABox -- systematic analyses of issues
in fly scans are conducted.  Interesting products of these analyses include
a general-purpose software-based fly-scan mechanism, a general way to design
undulator-monochromator fly scans, a sketch of how to practically implement
online tuning of fly-scan behaviours based on processing of the
data acquired, and many more.  Based on the results above, an
architectural discussion on $\geq 10$\,kHz fly scans is given.

\section{Introduction: an architectural overview of fly scans}\label{sec:panda}

The High Energy Photon Source (HEPS) \cite{jiao2018} is a 4th-generation
synchrotron radiation facility under construction at Beijing, China; it will
provide 14 beamlines to users (plus 1 internal test beamline) in its Phase I,
and can provide up to 90 beamlines in total in future phases.  At HEPS,
diverse requrements in the numerous kinds of beamline experiments at HEPS
need to be handled with often limited human resourses; high framerates and
data throughputs, mandated in many experiments by the small X-ray spots
and high brightness, also must be supported.  To meet these requirements, the
\prog{Mamba} software project \cite{liu2022, dong2022} was initiated at HEPS
based on \prog{Bluesky} \cite{allan2019}, with the goal of becoming a reliable,
flexible, performant and maintainable software framework for beamline
experiments in mind.  Since then, \prog{Mamba} has been actively developed:
pilot applications of it have been deployed or are being tested at the
4W1B, 3W1, 4B7A and 4W1A beamlines of the Beijing Synchrontron
Radiation Facility (BSRF), and the list is still growing.

\begin{figure}[htbp]\centering
\includegraphics[width = 0.9\textwidth]{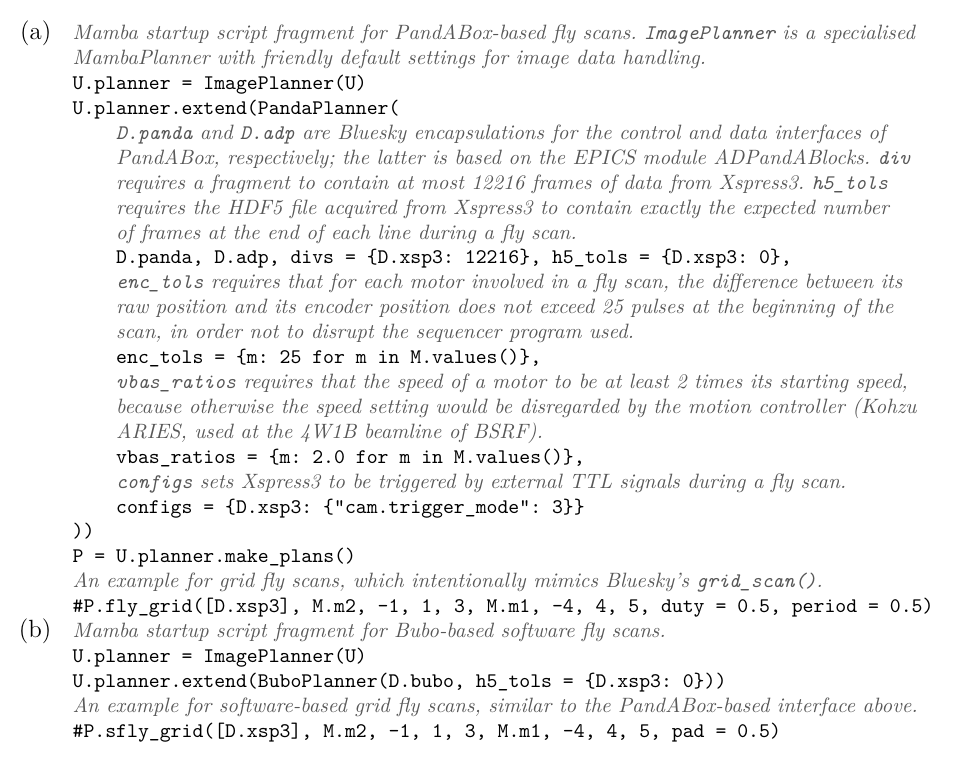}
\caption{Sketch of the usage for \prog{MambaPlanner}}\label{fig:planner}
\end{figure}

An extremely important part of the work above is the development of
\prog{Mamba}-based fly scans with PandABox \cite{zhang2017, christian2019}.
As was mentioned in our last fly-scan paper \cite{li2023b}, we were unsatisfied
with our previous command-line interface for fly scans, which we find too
bloated for both manual and automated use (the latter through command injection
\cite{liu2022}).  To solve this problem, we introduced what we call the
\prog{MambaPlanner} mechanism (\figref{planner}); apart from simply saving
repetitive inputs, it is also designed to perform various pre-scan and in-scan
correctness checks.  The check list now includes encoder errors, motor speeds
and (in-scan check for) HDF5 frame loss; this can be easily extended /
customised to support more sophisticated / site-specific requirements.
In the future, we also plan to add support in \prog{MambaPlanner} for mixing
detectors with different readying delays and trigger types (edge trigger or
level trigger), in order to allow users to easily combine detectors that fit
experiment-specific needs, saving them the burden to set PandABox parameters
for delaying and pulse shaping of trigger signals for individual detectors.

The above is a prime example for the notion of experiment parameter generators
(EPGs) in \prog{Mamba}, which was originally meant to save users the need to
input the same experiment information multiple times, as the name implies.
The idea has been generalised to simplify programming on multiple levels:
for users, so that they can focus on the methodology, instead of hairy
hardware/software details; for beamline operators, which only need to specify
essential hardware/software parameters, and not copy large amounts of code they
do not understand.  The pursuit of minimal yet expressive interfaces also urges
the programmer to construct minimal yet expressive implementations, so that
the burdens in both development and maintenance are minimised; on a deeper
level, these require the programmer to think about the inherent nature of the
problem in consideration, which often leads to novel insights.  This idea is
also the origin of this paper, which discusses what has been done and explores
what can be further done to systematically implement advanced fly scans; they
are based on what we believe to be the architectural essentials of fly scans.

\begin{figure}[htbp]\centering
\includegraphics[width = 0.4\textwidth]{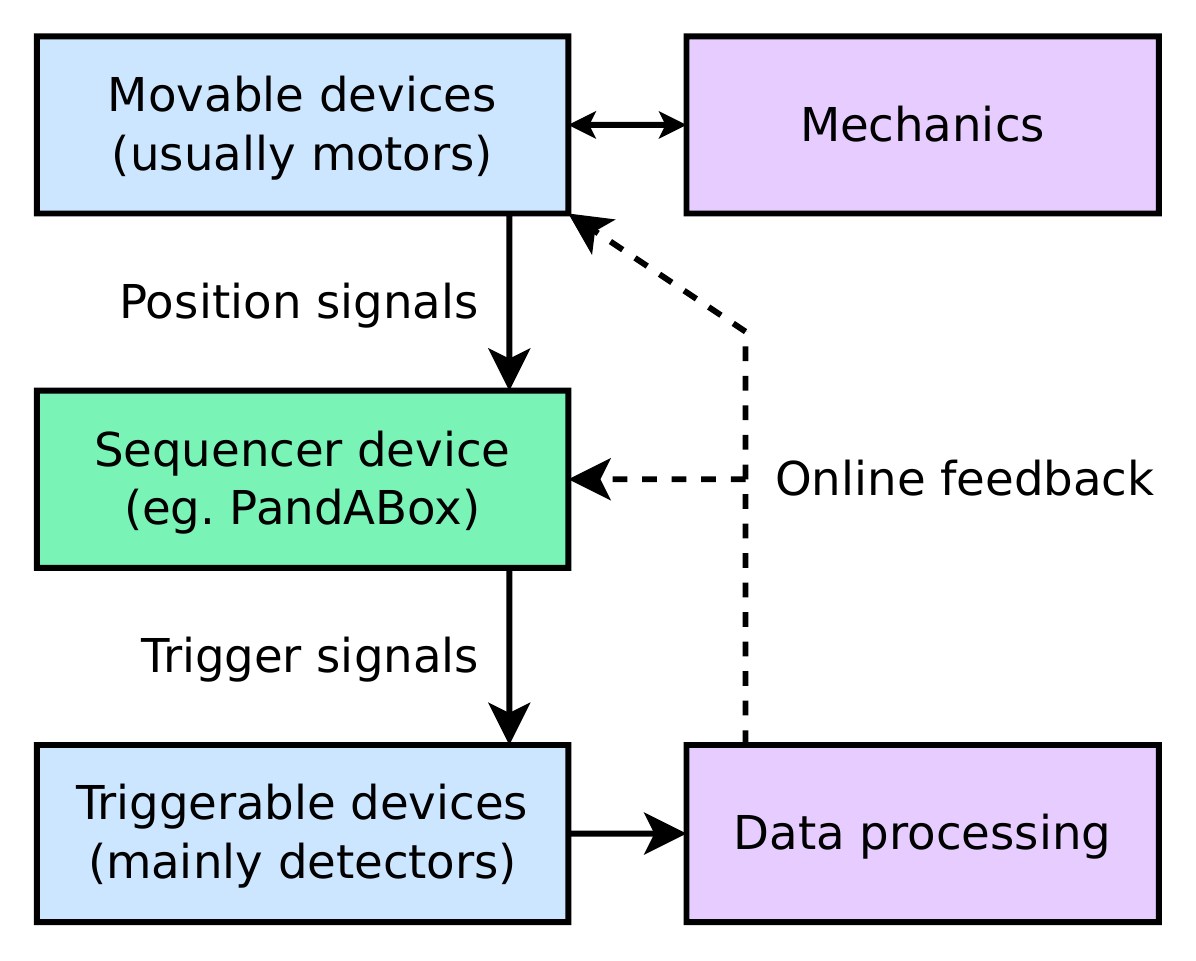}
\caption{The ``P(osition) and A(cquisition)'' architecture of fly scans}
\label{fig:panda}
\end{figure}

In our opinion, these essentials are nicely captured in the name ``PandA'' --
p(osition) and a(cquisition) (\figref{panda}): the former concerns movable
devices, most importantly motors, which need to produce position feedback; the
latter concerns devices that can be triggered to acquire data, most importantly
detectors.  Sequencer devices like PandABox accept position inputs and produce
trigger outputs according to configured sequencer programs; closely related to
motors and detectors are mechanics and data processing.  Thus a fly scan is
in principle feasible, as long as the motors (plus mechanics), the detectors
(plus data processing) and the sequencer program, considered as loosely coupled
subsystems, all behave correctly.  Correspondingly, in the following sections
of this paper, we will first discuss issues mostly related to motion control
and triggering sequences (\secref{motors}), then discuss issues more related
to detectors and other fields of interest in fly scans (\secref{dets}); and
after the reader is familiarised with our way of thinking, we present a
systematic discussion of $\geq 10$\,kHz fly scans (\secref{10khz}), which
are both state of the art \cite{deng2019, batey2022} and practically required
at multiple beamlines of HEPS.  In the rest of this section we briefly present
two examples which we directly benefit from the crude analysis above.

The first example is a software-based fly-scan mechanism that we call ``Bubo''.
Based on the observation above about sequencer devices, Bubo is designed to
be general-purpose and intentionally modeled to mimic fundamental behaviours
of PandABox where applicable; it also has a \prog{MambaPlanner} encapsulation
(\figref{planner}) highly similar to its PandABox-based counterpart.  Of course,
Bubo is not immune from the limitations intrinsic to software-based sequencers:
first of all, its performance is limited by the speed of software-based position
feedback and detector triggering, so attempts to exceed its performance upper
bound (which is quite low) can only result in sequencer disruption or frame loss
(the latter can be detected by \prog{MambaPlanner}).  It is also less accurate
than fly scans based on hardware like PandABox, as well as step scans, because
the position-to-trigger feedback time is much longer than its hardware-based
counterpart, and the accuracy of motor positions is much less guaranteed.
However, Bubo has actually never been expected to be a replacement for
devices like PandABox; except when explicitly noted, all ``fly scans''
discussed later in this paper mean PandABox-based scans.  Instead,
Bubo is more intended for experiments where the positions vary slowly --
especially many \emph{in situ} experiments in our own expectation;
when it gets employed in combination with the online tuning of fly-scan
behaviours (\cf\ \secref{dets}), exciting opportunities may emerge.

The second example is about angle-resolved photoemission spectroscopy (ARPES)
experiments, where a big fraction of the ``movable'' devices are electrostatic
lenses instead of real motors.  Like the environment parameters in \emph{in
situ} experiments, the parameters of electrostatic lenses can be controlled
to vary continuously with time, so we may well treat them as \emph{generalised
motors}.  Following this, we can define \emph{generalised motion control}: as
long as we can set the parameters and get their feedback, we can be said to
``control'' them.  As long as the trajectories of the parameters over time in
their configuration space can be controlled within a tolerable error limit,
and their feedback can be sent to devices like PandABox or Bubo, a fly scan
is in principle feasible.  Unlike the slow \emph{in situ} experiments mentioned
above, in ARPES experiments we expect the controlled parameters to vary as
fast as reasonable, so that the efficiency of experiments can be maximised.
At HEPS, we have already had some preliminary discussions with the
high-resolution nanoscale electronic structure spectroscopy
beamline (BC) on plans to actually approach this goal.

\section{Motor and sequencer issues in advanced fly scans}\label{sec:motors}

At HEPS, we currently use simple moves (done with \verb|grid_scan| from
\prog{Bluesky}) of motors when implementing constant-speed mapping, in
pursuit of simplicity in programming; another reason is that many motors
(or their EPICS IOCs) we use do not readily support profile moving.  However,
we also realise that in ultrafast mapping experiments where each line uses
very little time, the turnover time between lines in simple moves will become
a performance bottleneck because of the communication overhead involved.
So in this case, profile moving surely becomes mandatory; the experience
in ultrafast mapping will also prepare us for experiments involving more
general motion trajectories, especially the irregular trajectories used
in ptychography \cite{odstrcil2018, deng2019}.  Here we do not delve into
the details in profile moving, but instead discuss a somewhat simpler yet
useful enough application of it, which we call \emph{pseudo-step scans}.

\begin{figure}[htbp]\centering
\includegraphics[width = 0.4\textwidth]{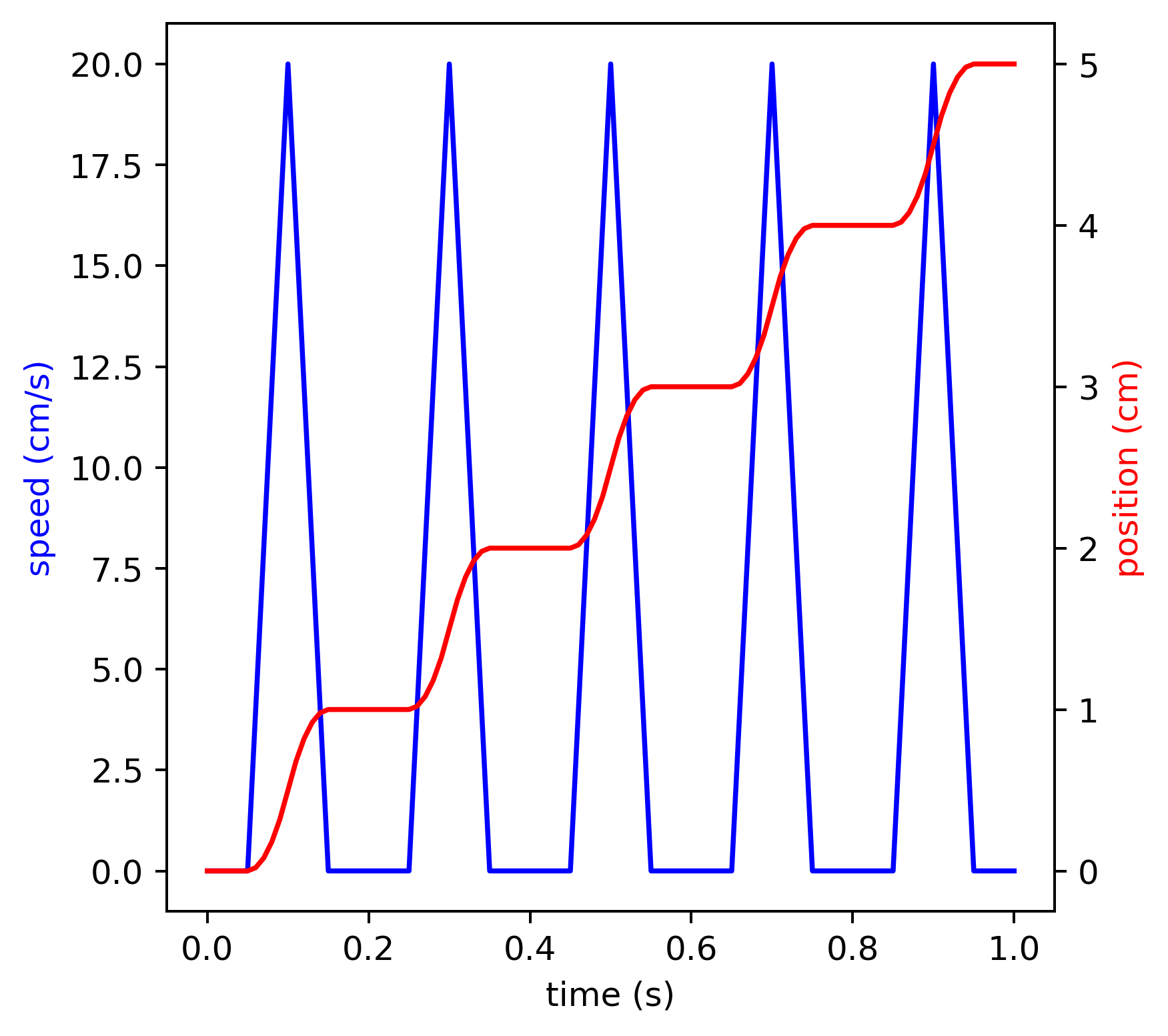}
\caption{Motion profile for a 5\,Hz pseudo-step scan}\label{fig:pseudo}
\end{figure}

Scanning transmission X-ray microscopy (STXM), a translation-based imaging
technique, is a main application at the hard X-ray imaging beamline (B7) of
HEPS.  Due to its translation-based nature, there is very little tolerance
for relative motion between the sample and the detector during the exposure
of an image frame, or otherwise obvious time-lapse effect would be observed.
Suppose we wanted to implement high-speed STXM with constant-speed fly scans,
the tolerance above, in conjunction with the lower bound on exposure time
imposed by the signal-to-noise ratio, would severely limit the motion speed,
resulting in an intolerably low duty ratio of exposure.  As this is a result
of the fact that most time is spent on slowly moving the sample, we can set
motion profiles that mimic regular step scans (\figref{pseudo}): just keep
the sample stationary when exposure is on, and quickly move the sample when
exposure is off.  In this way, while the exposure time is kept long enough,
the motion time can be minimised, so an optimal duty ratio can be achieved;
since the motion trajectories are still like those in grid scans, and the
exposure intervals and the darkness intervals are respectively constant-time,
sequencer programs from constant-speed fly scans can be easily adapted.

In our summary, the main obvious \emph{triggering strategies} in fly scans
are time-based and position-based: the former uses the moment when a specified
motor passes a certain checkpoint as a starting point, and then triggers a
series of exposures only based on time delays; the latter uses one checkpoint
to start each exposure, but the exposures times are the same to simplify data
processing.  There are quite a few other triggering strategies, but they can
usually be regarded as variants of the two strategies above; all the strategies
above are being tried at HEPS, in order to fully explore their potentials.
Time-based triggering is easy to implement with the built-in sequencer mechanism
in PandABox, and also seems to be the preferred strategy in \prog{pymalcolm}
\cite{basham2018}, the official middleware for PandABox (\cf\ the Python class
\verb|PandASeqTriggerPart| in \cite{cobb2022}).  When the deviation of the
actual motion profile from the preset profile (\ie\ the \emph{following error}
in motion control) is very small, time-based triggering is a particularly
preferable strategy for fly scans, but we also need to be care of the
\emph{clock drift} issue: when sufficiently long time has passed after
the latest checkpoint, the clock difference between two devices
(\eg\ PandABox and a motor) that do not share a clock may become
large enough to break assumptions in the triggering strategy.

In grid scans, whether implemented with simple moves or profile moving,
there is a checkpoint at the beginning of each line (which usually does not
take too much time), so clock drift is not a problem; in more complex scans,
\eg\ scans with irregular trajectories, it can make real troubles.  However,
by using \emph{scan fragmentation} \cite{li2023b}, we can deliberately split
long scans into short scans (``fragments'', \cf\ \figref{planner}), so that
the clock difference is zeroed before it is able to create any problem;
as long as the fragments are large enough, the performance penalty will be
relatively negligible.  Scan fragmentation also allows for the use of multiple
mechanically independent motors in a same fly scan: somehow make the beginning
of their motion hardware-synchronised (\eg\ with TTL signals), and then just
let each motor move on its own; their proper synchronisation is ensured by the
zeroing of clock drifts at the beginning of each fragment.  A major application
of this is fly scans involving insertion devices, \eg\ undulator-monochromator
fly scans, because we want to avoid connecting motors inside and outside
the accelerator to a same motion controller, as that would break the loose
coupling between the accelerator and the beamlines.  At HEPS, we have
confirmed that the motion controllers used for insertion devices support
motion synchronisation with hardware signals, and we will also ensure those
controllers on beamlines that get involved in these scans support it.

For fly scans with complex trajectories, time-based triggering may also
be the preferable strategy, as we just need to select some good checkpoints;
with position-based triggering, it would be more tricky to set the triggering
conditions, as we must avoid using the positions of motors that are too
slow, which may vary from one scan point to another.  Another way to do
position-based triggering is to transform the motor positions to some other
coordinates that are more readily usable as criteria for position comparison:
\eg\ transforming the positions in a spiral scan to approximated spiral path
lengths between the origin and the points in question.  This is non-trivial
with PandABox, but is fairly easy with devices like DeltaTau PowerPMAC; the
latter can accept motor encoder inputs from multiple (much more than PandABox's
4) ports, perform transformations and output the results using its encoder
output ports.  We additionally note that this is a general solution for
the problem of encoder processing, and one of PandABox's original purpose
was to tackle a special case of it \cite{zhang2017}; moreover, PowerPMAC's
support for a wide variety of encoders also makes it suitable as an
encoder converter for PandABox that extends the latter's supported list
of encoder types, even if used without any coordination transformation.

\section{Detectors and other issues in advanced fly scans}\label{sec:dets}

\begin{figure}[htbp]\centering
\includegraphics[width = 0.8\textwidth]{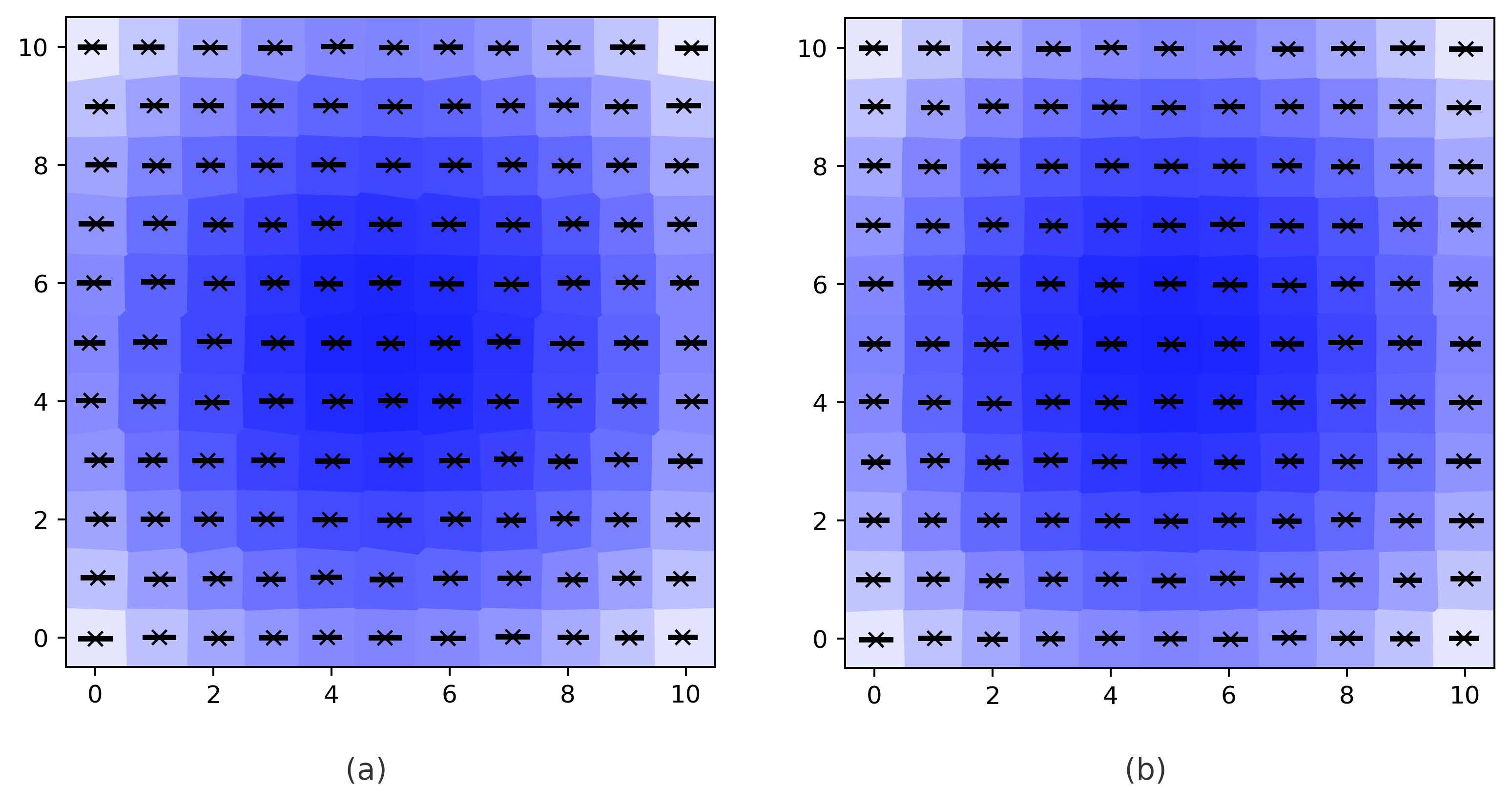}
\caption{%
	Voronoi visualisation of results from a hypothetical grid scan with
	a snaking motion trajectory and exaggeratedly obvious following errors:
	(a) using time-based triggering and (b) using position-based triggering.
	Crosses mark mean motor positions during the exposures, and bars
	mark maximal / minimal positions during the exposures; we also
	note that bars on the odd and even rows in (b) are respectively
	left-aligned and right-aligned because of position-based
	triggering and the snaking trajectory.%
}\label{fig:voronoi}
\end{figure}

In \secref{motors}, we have mentioned the issue of following errors, which is
also an issue with constant-speed fly scans; we have noted that time-based
triggering is particularly suitable when the following error is very small.
On the other hand, as was mentioned in our last fly-scan paper \cite{li2023b},
even with large following errors or even deliberately irregular motion
trajectories, we can plot the results as Voronoi diagrams (\figref{voronoi}).
We note that it may be observed from the figure that even with what we call
position-based triggering here, with obvious following errors Voronoi diagrams
are still necessary when doing visualisation.  Another choice is to use purely
position-based triggering: the end of each exposure is determined by the
moment the motor leaves the preset exposure interval; in this case, since
the exposure time is no longer constant, an extra normalisation step must be
performed for the data acquired.  A more important issue here we note is that
with irregular motion trajectories, we obviously cannot solely use minimal,
maximal and mean values of the motor positions, like those in \figref{voronoi},
to characterise the following errors.  Instead, for each frame of exposure we
must collect multiple samples of the motor positions (\eg\ $\geq 50$ samples
in \cite{deng2019}); this is doable with PandABox, but requires more complex
configurations than those where only one sample is needed for each frame.

A deeper issue with PandABox's position capture is its number of motor
encoder ports: 4 input and 4 outputs; since piezoelectric motors can send
their positions to the PandABox through the latter's ADC card, the number of
input channels for them is larger.  In complex experiments, \eg\ some at the
hard X-ray nanoprobe multimodal imaging beamline (B2) and hard X-ray coherent
scattering beamline (B4) of HEPS, the number of position channels that need
to be captured can exceed PandABox's limit.  Other than modifying PandABox
itself, another choice is directly reading positions from the motors, or
in other words treating them as \emph{0-dimensional detectors}: the motors
should be able to store the positions in a ring buffer, which can be retrieved
by the controlling computer through some kind of programming interface; this
is the case with many advanced motion controllers that support profile moving,
like those from DeltaTau, SmarAct and ACS.  On the computer side, we find the
current EPICS-based software infrastructure to be unsatisfactory in this aspect:
we have the \prog{areaDetector} framework for over 1-dimensional detectors,
whose design makes it non-trivial to adapt to 0-dimensional requirements.

We note that in terms of performance, \prog{areaDetector} also has significant
limitations: when writing to HDF5 files with its \emph{HDF5Plugin}, the
framerate is limited to roughly 4\,kHz, and the data throughput is limited to
roughly 500--600\,MB/s; in comparison, a single detector at the B7 beamline
of HEPS will be able to continuously produce data at 8.5\,GB/s, even only
in Phase I of HEPS.  The former, for instance, cannot fulfill the needs
in \secref{10khz} even for the capture of data from PandABox itself (if we
use the \prog{areaDetector} module \prog{ADPandABlocks} for this, \cf\ %
\figref{planner}).  At HEPS, we are developing a high-performance workalike of
\prog{areaDetector} that, in conjunction with software like \prog{Mamba Data
Worker} (\cf\ \secref{10khz}), can overcome these limitations, and are already
nearing the completion of its prototype.  Detectors themselves may also pose
problems: in addition to framerate / throughput limits, we would also like
to emphasise limitations on other hardware capabilities, like countrate limits.
Readout systems for silicon drift detectors (SDDs), like Xspress3 and Falcon
which are widely used in X-ray fluorescence (XRF) spectroscopy, have a countrate
limit of around 3--4\,Mcps; noticing that with the usual frame size (4096 bins),
with a 1\,kHz framerate each bin would receive about only one count on average.
We can see that in this situation, the signal-to-noise ratio is already
terrible; to make things even worse, this is when the readout system gets
highly saturated, so serious deadtime effects would further reduce the
signal-to-noise ratio.  It follows that with the development of ultrafast
scans, SDD readout systems with higher countrate limits will be
of great scientific and commercial values.

\begin{figure}[htbp]\centering
\includegraphics[width = 0.8\textwidth]{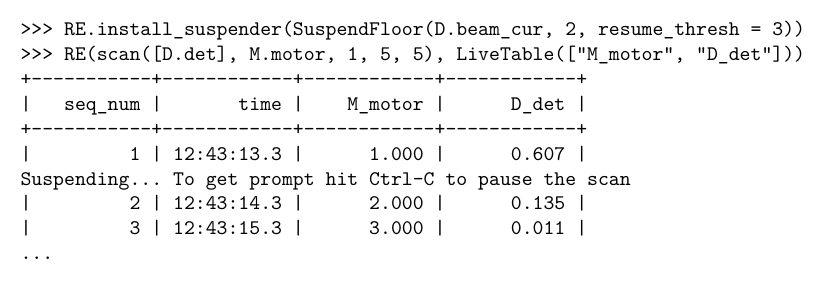}
\caption{%
	Example usage of \prog{Bluesky}'s ``suspender''
	functionality based on monitoring of the beam current%
}
\label{fig:suspend}
\end{figure}

In our \prog{Mamba} paper \cite{liu2022} we mentioned our goal of dynamic tuning
of the behaviours in fly scans based on online processing of data acquired
during the scans (\figref{panda}).  In order to be able to implement this,
the scans need to be dynamically tunable in the first place: the scan mechanism
needs to be able to accept new instructions which depend on the processing of
data acquired during the execution of old instructions.  A nice candidate for
this requirement is the double-buffer design in \prog{pymalcolm} \cite{dls2015},
which flip-flops between PandABox's two sequencer blocks, each of which can
hold up to 4096 sequencer instructions, so that in principle an endless stream
of instructions can be fed to PandABox.  A somewhat similar mechanism can be
implemented on the motor side to dynamically accept new motion instructions,
thus completing the main architectural elements necessary for the dynamic tuning
of fly scans.  Apart from tuning of regular fly-scan behaviours, automatic
pausing/resuming (\eg\ base on status monitoring of the accelerator) is also
a part of our goal; the automatic abortion of scans by \prog{MambaPlanner} upon
the detection of errors (\cf\ \secref{panda}) can be considered as a simplified
variant of this goal.  \prog{Bluesky}'s \verb|RunEngine| has a ``suspender''
functionality that automatically triggers pausing/resuming (\figref{suspend}),
but its pausing/resuming mechanism is insufficient for fly scans because of the
need to rewind scan points: it only supports rewinding up to one scan point, but
in fly scans more points need to be rewound because of the fast acquisition and
asynchronous processing of data.  The number of points to rewind should ideally
be deduced from processing of the data acquired, in order to rule out data
acquired when a pause is already necessary but the pause action has not
yet been fired.  The addition of \verb|RunEngine| support for full-fledged
pausing/resuming of fly scans is already on our middle-term development roadmap.

\section{$\geq 10$\,kHz fly scans: an architectural analysis}\label{sec:10khz}

Among the 15 beamlines at HEPS (Phase I), there are at least 2 beamlines
that require fly scans with 10\,kHz or higher framerates: the B4 beamline in
its coherent diffractive imaging (CDI) and X-ray photon correlation spectroscopy
(XPCS) experiments, and the BC beamline in its 2-dimensional real-space current
mapping experiments; of course, the list is going to grow with time.  As fly
scans are a collaboration between several loosely coupled subsystems, we analyse
the potential issues by the subsystems (\figref{panda}): the sequencer device,
detectors, data processing, mechanics / motion control (in the order which we
follow in this section).  So first of all, as devices like PandABox are based on
FPGAs with clock frequencies at least on the MHz level, with proper hardware and
firmware designs that allow users to fully exploit their clock cycles, these
devices themselves should not become the bottleneck by themselves.  PandABox
has a well-designed interface that fully utilises its clock cycles, but its
number of motor encoder input/output ports and limited support for encoder
types may become the limiting factors in certain experiments (although not
necessary $\geq 10$\,kHz).  This is however solvable, and our solutions
have been proposed in Section \ref{sec:motors}--\ref{sec:dets}.

The next subsystem is detectors, including data readout.  In order to do
$\geq 10$\,kHz fly scans, the detectors themselves obviously need to be able to
accept external triggers, and offer sufficient framerates with acceptable data
quality; furthermore, the readout software should not result in excessive
downgrades in framerates or data throughputs.  In \secref{dets}, we have
discussed the performance issue with EPICS's \prog{areaDetector} as an example
for issues with the readout software; in the same section, the countrate limit
of current SDD readout systems is given as a real-world example for the data
quality issue.  Another data quality issue we can imagine is the possibility
of distortion in signals fed to PandABox through the latter's ADC card, which
is a composition of distortion from the ADC card, any preamplifier and other
components involved in transmission of the signals.  There is the possibility
that the distortion was smoothed out in fly scans of lower frequencies, but
got revealed in $\geq 10$\,kHz scans; should this really happen, it would need
to be reduced by improved experiment design and handled in data processing.
Data processing is also the subsystem that comes after detectors; in our
summary it includes transmission, storage and computational processing.
The main challenge in it is the design and implementation of a reliable,
flexible, performant and maintainable hardware/software system
architecture that does the things above.  At HEPS, \prog{Mamba Data Worker}
(\prog{MDW}) \cite{li2023a} is the software framework that lies at the heart
of this architecture; as a high-throughput data orchestration and processing
system, it is also a core component of the \prog{Mamba} framework.

\begin{figure}[htbp]\centering
\includegraphics[width = 0.4\textwidth]{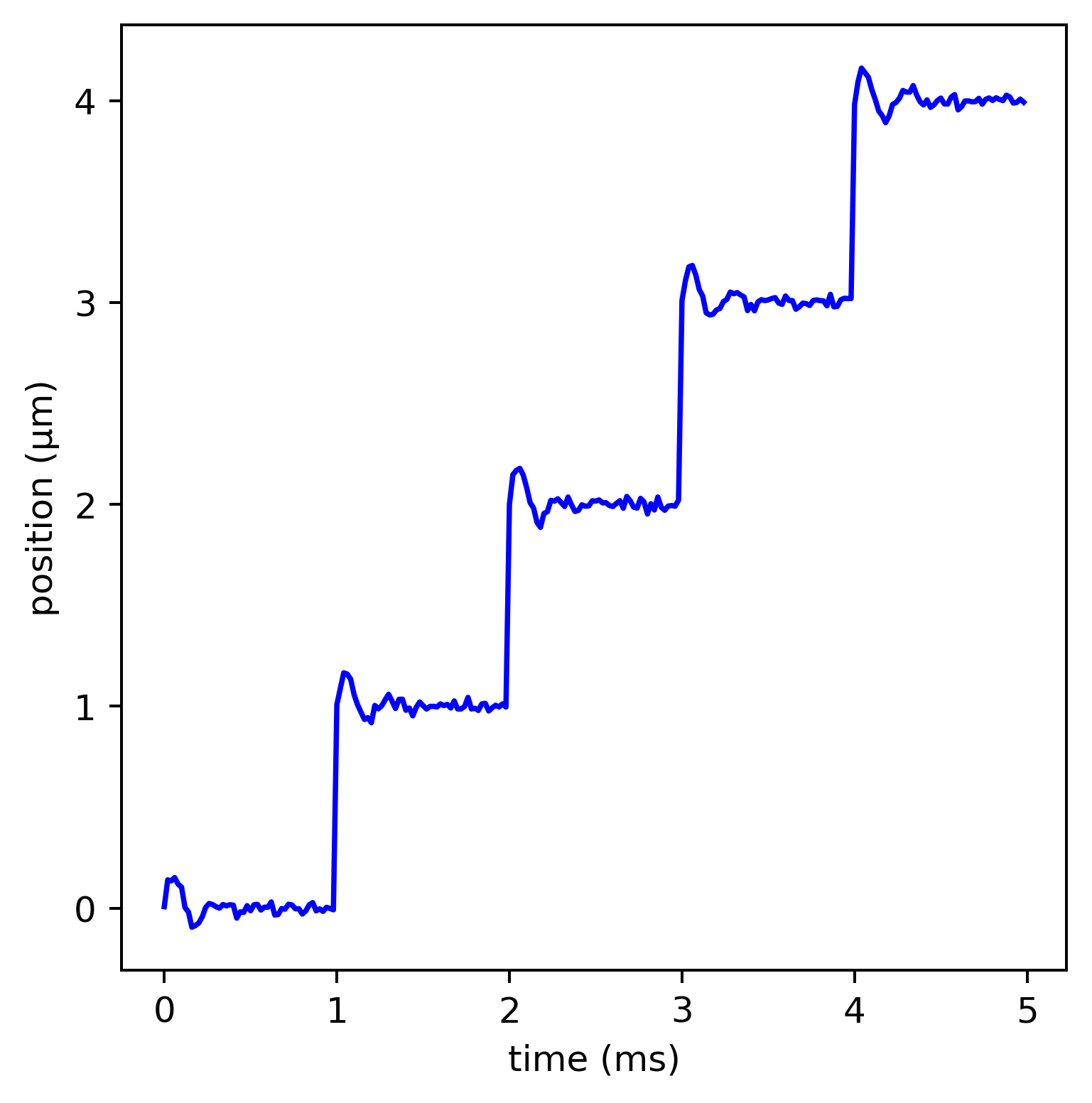}
\caption{%
	Simulation of a stepping motor's motion profile measured
	with a laser interferometer under a high frequency%
}\label{fig:stepping}
\end{figure}

The final and most complex subsystem is mechanics and motion control.
Mechanics has already been a biggest challenge on multimodal beamlines,
\eg\ the B2 beamline of HEPS, because of the versatility required for the
mechanical system involved.  At the B4 beamline of HEPS, significant challenge
has also been observed due to the mechanical properties required to allow for
$\geq 10$\,kHz fly scans with satisfactory quality.  On the motion control side,
aside from the usual difficulties in terms of control, we note that electronic
position feedback may not fully reflect the amplitudes of the sample's vibration
in locations relatively far from where the position probes monitor; this may
become a problem in scans with irregular motion trajectories, and pseudo-step
scans (\cf\ \secref{motors}, although probably not $\geq 10$\,kHz).  Another
issue is about the nature of stepping motors: although their motion profiles
may seem smooth when observed under low frequencies, with $\geq 10$\,kHz
sampling it will be revealed that the profiles are more like step functions
(\figref{stepping}).  This is somewhat similar to the signal distortion
issue analysed above, but unlike the issue above, it is not just theoretical.
It renders stepping motors unfit for the ``flying axes'' in $\geq 10$\,kHz
scans, as we really want their positions to smoothly vary with time when
observed under the scan frequency; instead, servo or piezoelectric
motors can be used, as they do not have this limitation.

\section{Conclusion}

Fly scans can be regarded as a collaboration between a few loosely coupled
subsystems: motors / mechanics, detectors / data processing, and sequencer
devices like PandABox.  Based on this observation, Bubo, a general-purpose
software-based fly-scan mechanism, is introduced; the notion of generalised
motion control is introduced, which for instance can facilitate the
implementation of fly scans in ARPES experiments.  The idea of pseudo-step
scans is presented to resolve the motion speed issue in translation-based
imaging.  The clock drift issue in long fly scans with time-based triggering
is noted, and scan fragmentation is proposed as a general solution to it;
a major application of it is the implementation of undulator-monochromator
fly scans.  DeltaTau PowerPMAC is proposed as a general solution for the
problem of encoder processing / coordination transformation, and it can
also be used as an encoder converter for PandABox.  In addition to capturing
motor positions with PandABox, it is also possible to directly capture the
positions from the motors themselves, treating them as 0-dimensional detectors;
the corresponding deficiencies of EPICS \prog{areaDetector} are noted, and we
are developing a workalike for it that overcomes these deficiencies.  Apart
from limits in framerates and throughputs of detectors, we also note that other
limitations, like the countrate limit of current SDD readout systems, may also
limit the speed of fly scans.  By using strategies like the double-buffer
design in \prog{pymalcolm}, in principle endless streams of sequencer/motion
instructions can be fed to PandABox/motors, paving way to the online tuning
of fly scans based on processing of the data acquired; the problems with
attempts to implement automatic pausing/resuming of fly scans are also
discussed.  An architectural analysis is given for $\geq 10$\,kHz fly
scans.  The potential issue of ADC signal distortion in high-frequency
data captures is discussed; similarly, it is noted that stepping motors
are unsuitable for the ``flying axes'' in $\geq 10$\,kHz fly scans.

\section*{Statements and declarations}

\paragraph{Acknowledgements:}
We would like to thank all beamlines at BSRF and HEPS, especially those
explicitly mentioned in this paper, for fruitful discussions about fly scans.

\paragraph{Funding:}
This work was supported by the Young Scientists Fund of the National Natural
Science Foundation of China (Grants Nos.\ 12005253, 12205328) and the
Technological Innovation Program of Institute of High Energy Physics
of Chinese Academy of Sciences (Grant No.\ E25455U210).

\paragraph{Data availability:}
The source code of Bubo has been released as a part
of a fully open-source edition of \prog{Mamba} available at
\url{https://github.com/CasperVector/mamba-ose}; it depends on
currently HEPS-specific patches for \emph{Bluesky} components available at
\url{https://github.com/CasperVector/ihep-pkg-ose/tree/master/misc/pybuild}.

\bibliography{art7}
\end{document}